\documentstyle[sprocl,epsf,rotate]{article}

\def\Journal#1#2#3#4{{#1} {\bf #2}, #3 (#4)}

\def\ApJ{\em Astrophys. J.}

\def\NPB{{\em Nucl. Phys.} B}
\def\PLB{{\em Phys. Lett.}  B}
\def\PRL{\em Phys. Rev. Lett.}
\def\PRD{{\em Phys. Rev.} D}
\def\RMP{\em Rev. Mod. Phys.}

\def\be{\begin{equation}}
\def\ee{\end{equation}}
\def\bea{\begin{eqnarray}}
\def\eea{\end{eqnarray}}

\begin{document}
\pagestyle{empty}

\title{SELECTED TOPICS IN NEUTRINO ASTROPHYSICS}

\author{ A.B. BALANTEKIN \footnote{ Electronic Address:
baha@nucth.physics.wisc.edu}}

\address{University of Wisconsin, Department of Physics, \\ Madison,
WI 53706, USA}

\maketitle\abstracts{ This review covers a subset of the astrophysical
phenomena where neutrinos play a significant role and where the
underlying microphysics is nuclear physics.  The current status of the
solar neutrino problem, atmospheric neutrino experiments, the role of
neutrinos in determining the dynamics of type-II supernovae, and
recent developments in exploring neutrino propagation in stochastic
media are reviewed.}

\section{Introduction}

\indent
 
Neutrinos are perhaps the most interesting particles that we {\it
know} to exist. (Of course there are many other interesting particles
such as axions, supersymmetric partners, etc., that we {\it wish} to
exist). It is gratifying to see that Federal Agencies in the U.S. and
their counterparts around the world share the rational exuberance of
neutrino physicists by funding a number of potentially very
interesting experiments.

This review is intended for non-experts in neutrino physics and 
astrophysics. Topics covered here are chosen from those phenomena
where the underlying microphysics is nuclear physics. The current status
of the solar neutrino problem, atmospheric neutrino experiments, the
role of neutrinos in determining the dynamics of type-II,
core-collapse supernovae, and recent developments in exploring
neutrino propagation in stochastic media are reviewed. One should
emphasize that this is not intended to be a complete list. Other very
interesting astrophysical and cosmological phenomena in which
neutrinos play a role, such as
primordial nucleosynthesis, the possibility of hot dark matter, and
possible connection between neutrinos and pulsar proper motions are
excluded from this review. Similarly, high-energy neutrino
astrophysics, such as using neutrinos to probe active galactic nuclei,
are omitted.

Neutrino physics and astrophysics are rapidly developing areas. It
would be very difficult for any printed review to remain current for a
sufficiently long time. To keep up to date on the current
developments would require a dynamic medium such as the World Wide
Web. Indeed, there are a number of sites devoted to neutrino
physics. To keep up with the experimental developments some excellent
sites to consult include ``The Neutrino Oscillation Industry'' page
\cite{industry} at Argonne National Laboratory and ``The Ultimate
Neutrino Page'' \cite{ultimate} at Helsinki. For theoretical results
recommended sites include Bahcall's homepage\cite{bahcallpage} at the
Princeton Institute for Advanced Study and the ``Implications of Solar
Neutrino Experiments'' page \cite{implications} at the University of
Pennsylvania.

\section{Neutrino Oscillations}

\indent

In the minimal $SU(2) \times U(1)$ Standard Model neutrinos are taken
to be massless (i.e., the neutrino field is left-handed). However, the
Standard Model is an effective field theory \cite{lepage}, applicable
up to a certain energy scale beyond which new physics occurs. A
massive neutrino would be one of the indications of such new physics
and the value of the neutrino mass would be related to the scale of
new physics \cite{valle}. For example, in the Grand Unified theories
(GUTs) the right-handed neutrino sits in a singlet representation of
the weak $SU(2)$ taken to be a subgroup of the fundamental
representation of the grand-unifying gauge group. Since both helicity
components are present in the theory, the spontaneous symmetry
breaking mechanism yields a mass term for neutrinos in the same way it
does for other leptons. Since both helicity components of the {\it
neutrino} contribute to this mass term, it is known as the ``Dirac
mass''. For an electrically neutral particle such as the neutrino it is
possible to introduce another mass term by combining the left-handed
neutrino field with the right-handed {\it antineutrino} field, which
both appear as weak isodoublets in the Standard Model. Such a mass
term is known as the ``Majorana mass''. A Majorana mass term leads to
a breakdown in lepton number conservation. Neutrino masses can be
introduced in GUTs via the See-Saw mechanism \cite{csaw}. The See-Saw
mechanism produces light neutrino masses of the order of $m_{\nu} \sim
M^2_D/M_M$, where $M_D$ is a Dirac neutrino mass, and $M_M$ is a
Majorana neutrino mass which is a weak isosinglet. Since the latter is
associated with a new physics scale which is typically much larger
than the electroweak scale associated with the former, this mechanism
naturally produces very small neutrino masses.

In general the neutrino weak eigenstates that are observed in nuclear
decays are not mass eigenstates. In these lectures we consider
two-neutrino mixing for definiteness and assume that there is the
usual unitary transformation between the flavor eigenstates (e.g.,
$\Psi_e$ and $\Psi_{\mu}$) and the mass eigenstates ($\Psi_2$ and
$\Psi_1$):
\begin{equation}
\left[\begin{array}{cc} \Psi_e \\ \\ \Psi_{\mu}\end{array}\right] =
\left[\begin{array}{cc} \cos{\theta_v} & \sin{\theta_v}\\ \\
-\sin{\theta_v} & \cos{\theta_v}
\end{array}\right]
\left[\begin{array}{cc} \Psi_1 \\ \\ \Psi_2 \end{array}\right]\,,
\end{equation}
where $\theta_v$ is the vacuum mixing angle between the two flavors. A
generalization to the case of three flavors is straightforward. 

\subsection{Vacuum Oscillations}

\indent

Since the electron neutrino produced, for example, in a nuclear decay
is a linear combination of the mass eigenstates and these mass eigenstates
propagate with different phases, there is a probability that the other
weak eigenstate will appear after some distance $L$. The evolution of
the flavor eigenstates is governed by the equation
\begin{equation}
i\hbar \frac{\partial}{\partial t} \left[\begin{array}{cc} \Psi_e(t)
\\ \\ \Psi_{\mu}(t)\end{array}\right] = \frac{1}{4 E}
\left[\begin{array}{cc} - \delta m^2 \cos{2\theta_v} & \delta m^2
\sin{2\theta_v}\\ \\ \delta m^2 \sin{2\theta_v} &  \delta m^2
\cos{2\theta_v}
\end{array}\right]
\left[\begin{array}{cc} \Psi_e(t) \\ \\ \Psi_{\mu}(t)
\end{array}\right]\,,
\end{equation}
where $E$ is the neutrino energy, and $\delta m^2 = m_2^2-m_1^2$ ($m_2 >
m_1$).  In this equation a term proportional to the identity has been
dropped since it does not contribute to the relative phase between the
$\nu_e$ and $\nu_{\mu}$ components.  The appearance probability of the
other flavor is then given by \be P = \sin^2 2\theta_v \sin^2 (1.27
\delta m^2 L /E), \ee In the above equation $\delta m^2$ is measured
in eV$^2$ and $L/E$ in m/MeV. Neutrino oscillation experiments are
somewhat arbitrarily divided into two classes: short-baseline and
long-baseline. As Eq. (3) indicates, the longer is the baseline, $L$,
the more sensitive the experiment is to smaller values of $\delta
m^2$.

So far no group, with one possible exception, has reported a positive
result in either the appearance nor in the disappearance
experiments. The Liquid Scintillator Neutrino Detector (LSND)
collaboration at Los Alamos National Laboratory reported a significant
``oscillation-like'' excess \cite{lsnd}. This detector sits 30 meters
away from the beam dump at Los Alamos Meson Physics Facility
(LAMPF). The experimental apparatus is designed to produce a beam of
muon antineutrinos with as little contamination as possible from the 
electron antineutrinos. If these muon antineutrinos oscillate into
electron antineutrinos, such formed electron antineutrinos would interact
with protons in the detector, creating a positron and a neutron. This
neutron, after some time, binds with a proton to form a deuteron,
giving a photon with a characteristic energy of 2.2 MeV. The
experiment observes these photons as well as the positron's Cerenkov
track. When they identify both signatures together, the LSND group
first reported seeing nine events versus an expected background of two
events coming from the electron antineutrinos from sources other than
the muon antineutrino oscillation. This signal received a lot of
attention and a fair amount of criticism. Indeed, a dissenting member
of the LSND team has performed a data analysis of his own which finds
no positive signal above the expected background \cite{hill}. The LSND
result is unlikely to be a statistical fluctuation. However, the KARMEN
collaboration, carrying out a similar (but not identical) experiment
at Rutherford Laboratory in England, reported no evidence for neutrino
oscillations in a parameter space which largely overlaps with that of
LSND \cite{karmen}. Later the LSND collaboration reported 22
events. \cite{newlsnd} The LSND collaboration is currently analyzing
$\mu$ decay in flight, which will have different systematics and
backgrounds from the decay at rest analysis.

\subsection{Matter-Enhanced Oscillations}

\indent

Dense matter can significantly amplify neutrino oscillations. The mass
energy relation for free particles \be E^2 = p^2 + m^2 \ee is modified
in matter as a coherently forward-scattered neutrino acquires a potential 
via its interactions with the background particles
 \be (E-\phi)^2
= p^2 + m^2.  \ee In general this background potential is a Lorentz
four-vector. In Eq. (5) only the scalar component is included. The
contribution of the three vector component averages to zero for a
non-polarized medium or one without large scale currents. This scalar
potential is proportional to the strength of the weak interaction,
$G_F$. Hence ignoring terms proportional to $\phi^2$, Eq. (5) implies
an effective mass \be m_{\rm eff}^2 = m^2 + 2 E \phi.  \ee Including
this additional mass term the vacuum flavor evolution equation of
Eq. (2) is modified to
\begin{equation}
i\hbar \frac{\partial}{\partial t} \left[\begin{array}{cc} \Psi_e(t)
\\ \\ \Psi_{\mu}(t)\end{array}\right] = \frac{1}{4 E}
\left[\begin{array}{cc} A - \delta m^2 \cos{2\theta_v} & \delta m^2
\sin{2\theta_v}\\ \\ \delta m^2 \sin{2\theta_v} & -A + \delta m^2
\cos{\theta_v}
\end{array}\right]
\left[\begin{array}{cc} \Psi_e(t) \\ \\ \Psi_{\mu}(t)
\end{array}\right]\,. 
\end{equation}
In Eq. (7) $A$ depicts the neutral electronic medium correction to the
neutrino mass coming from the charged weak current, first calculated
by Wolfenstein \cite{wolf}
\begin{equation}
A = 2 \sqrt{2}\ G_F N_e(t) E\,,
\end{equation}
where $N_e(t)$ is the number density of electrons in the medium. There
is also a contribution coming from neutral weak current. However this
contribution is the same for all neutrino flavors and only effects the
overall phase. Eq. (7) also illustrates the reason behind the
enhancement of the neutrino oscillations in matter, namely level
crossing. For a given medium such as the Sun, where the density
changes by several orders of magnitude, the diagonal terms in Eq. (7)
vanish for a variety of neutrino parameters:
\be
 2 \sqrt{2}\ G_F N_e(t) E = \delta m^2 \cos{2\theta_v},
\ee
causing an efficient transformation between neutrino flavors without
any fine-tuning of the neutrino parameters. This behavior, first
noticed by Mikheyev and Smirnov \cite{ms}, is known as the
Mikheyev-Smirnov-Wolfenstein (MSW) resonance \cite{othermsw}.

Most of the salient features of matter-enhanced neutrino oscillations
can easily be addressed in the adiabatic basis \cite{adiabatic}. The
adiabatic basis is obtained by instantaneously diagonalizing the
Hamiltonian in Eq. (7). It is given by
\begin{equation}
\left[\begin{array}{cc} \Psi_i(t) \\ \\ \Psi_{ii}(t) \end{array}\right] =
\left[\begin{array}{cc} \cos{\theta(t)} & -\sin{\theta(t)} \\ \\
\sin{\theta(t)} & \cos{\theta(t)}
\end{array}\right]
\left[\begin{array}{cc} \Psi_e(t) \\ \\ \Psi_{\mu}(t)
\end{array}\right]\,,
\end{equation}
where the matter mixing angles are defined via 
\begin{equation}
\sin{2\theta(t)} = \frac{\sin{2\theta}_v}{\sqrt{\sin^2{2\theta}
 + \varphi^2(t)}}
\end{equation}
and
\begin{equation}
\cos{2\theta(t)} =  \frac{\varphi(t)}{\sqrt{\sin^2{2\theta} +
\varphi^2(t)}}\,,
\end{equation}
where 
\be
\varphi(t) = \frac{A}{\delta m^2} - \cos{2\theta}_v.
\ee
The matter angle defined this way changes from $\pi/2$ at infinite
density to $\theta_v$ in vacuum.  At the MSW resonance it takes the
value $\theta = \pi/4$. In this basis the evolution equation becomes
\begin{equation}
i\frac{\partial}{\partial t} \left[\begin{array}{cc} \Psi_i(t) \\ \\
\Psi_{ii}(t) \end{array}\right] = \left[\begin{array}{cc} - 
\sqrt{\sin^2{2\theta_v}
+ \varphi^2(t)} & -i \theta'(t) \\ \\ i  \theta'(t) &
\sqrt{\sin^2{2\theta_v} + \varphi^2(t)} 
\end{array}\right]
\left[\begin{array}{cc} \Psi_i(t) \\ \\ \Psi_{ii}(t) \end{array}\right]\,,
\end{equation}
where the prime denotes derivative with respect to the argument. 
The adiabaticity parameter is defined as
\begin{equation}
\gamma(t) =
\left|\frac{\sqrt{\sin^2{2\theta} + 
\varphi^2(t)}}{i \theta'(t)}\right|\,.
\end{equation}
When this parameter $\gamma(t)$ is large (the ``adiabatic limit''), we
can neglect the off-diagonal terms.  All nonadiabatic behavior, i.e.,
hopping from one mass eigenstate to the other, takes place in the 
neighborhood of the MSW resonance, $A = \delta m^2 \cos{2\theta}$.
The resulting electron neutrino survival probability averaged over the
detector location can be written as 
\be
P(\nu_e \rightarrow \nu_e) = \frac{1}{2} \left[1 + (1 - 2P_{hop})
\langle\cos{2\theta_i}\rangle_{src} \cos{2\theta_v}\right]\,, \ee 
where $\langle \ldots \rangle_{src}$ indicates averaging over the
initial source terms. It is possible to solve Eq. (7)
semiclassically to obtain the hopping probability \cite{beacom}
\be
P_{hop} = 
\exp\left(-i \frac{\delta m^2}{2 E}
\int^{t_0^*}_{t_0}
{dt \left[{\cal V}(t)\right]^{1/2}}\right),
\ee
where $t_0$ and $t_0^*$ are the turning points of the integrand and 
\be 
{\cal V}(t) = 
\left[\frac{2 \sqrt{2} G_F E N_e(t)}{\delta m^2}\right]^2
- 2\cos{2\theta_v}\left[\frac{2\sqrt{2} G_F E N_e(t)}{\delta m^2}\right] 
+ 1 .
\ee
The approximation leading to this expression is excellent in the
adiabatic regime and up to the extreme non-adiabatic limit. In the
latter limit logarithmic perturbation theory provides a useful
approach \cite{fricke}.

\section{Solar Neutrinos}

\subsection{Background Information}

\indent

The Sun is a main sequence hydrogen-burning star.  The energy that the
sun emits is released from nuclear fusion reactions among light
elements taking place near the center of the Sun.  The combined effect
of these reactions is \be 4p+ 2e^- \rightarrow ^4He + 2 \nu_e, \ee
with an energy release of 26.73 MeV.  If the stars are formed only of
hydrogen and helium, then nuclear reactions proceed via direct fusion
reactions between light elements.  The nuclear reactions in this
so-called pp chain are presented in Table 1.  If the star is formed
from a gas with an initial admixture of carbon, nitrogen, and oxygen,
then these heavier nuclei may serve as catalysts for the reaction in
Eq. (19), without themselves getting destroyed. The series of nuclear
reactions achieving this scenario is called the CNO-cycle.

\begin{table}[t]
\caption{
The pp chain nuclear reactions taking place in the Sun.}  \vspace{8pt}
\vspace{8pt} \centering
\begin{tabular}{|c|c|c|}
\hline {\em Reaction} & {\em Term. (\%)} & {\em $\nu$ energy (MeV)} \\
\hline \hline & & \\ \indent $p+p \rightarrow $D$ + e^+ + \nu_e$ &
99.96 & $\leq 0.420$ \\ \indent $p+e^-+p \rightarrow $D$ + \nu_e$ &
0.44 & \\ \hline \indent D$+p \rightarrow ^3$He$ + \gamma$ & 100& \\
\hline \indent $^3$He$ +^3$He$ \rightarrow ^4$He$ + 2p$ & 85& \\
\indent $^3$He$ +^4$He$ \rightarrow ^7$Be$ + \gamma$ & 15 & \\ \hline
\indent $e^-+ ^7$Be$ \rightarrow ^7$Li$ + \nu_e$ &15 & 0.861
(90\%),0.383 (10\%) \\ \indent $p+^7$Li$ \rightarrow 2 ^4$He & & \\
\hline \indent $p+^7$Be$ \rightarrow ^8$B$ +\gamma$ & 0.02 &\\ \indent
$^8$B$ \rightarrow ^8$Be$^* +e^+ +\nu_e$ & & $<15$\\ \indent $^8$Be$^*
\rightarrow 2 ^4$He & & \\ \hline \indent $^3$He$ + p \rightarrow
^4$He$ + e^+ + \nu_e$ & $4\times 10^{-6}$ & 18.8 \\ \hline
\end{tabular}
\end{table} 

To understand the evolution of the abundances of nuclear species is a
very instructive exercise.  The initial reaction network for the pp
chain in the Sun can be written as    
\begin{eqnarray*}
{d[H] \over dt} &=& -2 \lambda_{11} {[H]^2 \over 2} - \lambda_{12}
[H][D] + 2 \lambda_{33} {[^3He]^2 \over 2} - \lambda_{17} [H] [^7Be]
\\ & & \,\,\, -\lambda_{17}' [H] [^7Li] \\ 
{d[D] \over dt} &=& \lambda_{11} {[H]^2
\over 2}  - \lambda_{12} [H][D] \\ {d[^3He] \over dt} &=&
\lambda_{12}[H][D] - 2 \lambda_{33} {[^3He]^2 \over 2} - \lambda_{34}
[^3He] [^4He] \\ {d[^4He] \over dt} &=& \lambda_{33} {[^3He]^2 \over
2} - \lambda_{34} [^3He] [^4He] + 2 \lambda_{17} [H] [^7Be] + 2
\lambda_{17}' [H] [^7Li] \\ {d[^7Be] \over dt} &=& \lambda_{34} [^3He]
[^4He] - \lambda_{17} [H] [^7Be] -  \lambda_{e7} [e] [^7Be] \\
{d[^7Li] \over dt} &=& \lambda_{e7} [e] [^7Be] -\lambda_{17}' [H]
[^7Li]. \\
\end{eqnarray*}
In these equations the notation $[A]$ denotes the abundance of the
species $A$ and $\lambda_{AA'}$ represents the rate per pair of the
fusion reaction where the initial nuclides are $A$ and $A'$: \be
\lambda = \left( { 8 \over \pi \mu (kT)^3} \right)^{1/2} f_0
\int_0^{\infty} dE E \sigma(E) \exp{(-E/kT)}, \ee where $\mu$ is the
reduced mass of the system, and the factor $f_0$ takes into account
the screening of the bare nuclear charges by the electrons in the
solar plasma. The rate for the reaction $A + A' \rightarrow X$ is
given by \be r_{AA'}= \frac{[A][A'] \lambda_{AA'}}{1+\delta_{AA'}}.
\ee Determination of the cross sections and the screening factors is
where nuclear physics input to solar models is needed. A recent
comprehensive study of solar fusion rates is given in Ref. 18.

The lifetime for a nucleus of $A$ in the presence of the nucleus $A'$
is given by 
\be
\tau_A = \frac{1}{[A']\lambda_{AA'}}.
\ee
The deuterium lifetime in the Sun is extremely short, measured in
seconds. Hence deuterium can always be assumed in
equilibrium. Setting  ${d[D]/dt}=0$  in the previous set of
equations we obtain the
reaction network after the deuterium equilibration: 
\begin{eqnarray*}
{d[H] \over dt} &=& -3 \lambda_{11} {[H]^2 \over 2} + 2 \lambda_{33}
{[^3He]^2 \over 2} - \lambda_{17} [H] [^7Be]
-\lambda_{17}' [H] [^7Li] \\ {d[^3He] \over dt} &=& \lambda_{11}
{[H]^2 \over 2} - 2 \lambda_{33} {[^3He]^2 \over 2} - \lambda_{34}
[^3He] [^4He] \\ {d[^4He] \over dt} &=& \lambda_{33} {[^3He]^2 \over
2} - \lambda_{34} [^3He] [^4He] + 2 \lambda_{17} [H] [^7Be] + 2
\lambda_{17}' [H] [^7Li] \\ {d[^7Be] \over dt} &=& \lambda_{34} [^3He]
[^4He] - \lambda_{17} [H] [^7Be] -  \lambda_{e7} [e] [^7Be] \\
{d[^7Li] \over dt} &=& \lambda_{e7} [e] [^7Be] -\lambda_{17}' [H]
[^7Li]. \\
\end{eqnarray*}
Similarly $Li$ and $Be$ lifetimes are of the order of years in a
typical star.  After these nuclei reach equilibrium 
(${d([^7Be]+[^7Li])/dt}=0$) their abundances
remain proportional to the $^3He$ abundance. After the $Li$ and $Be$
equilibrium is reached the  reaction network takes the form 
\begin{eqnarray*}
{d[H] \over dt} &=& -3 \lambda_{11} {[H]^2 \over 2} + 2 \lambda_{33}
{[^3He]^2 \over 2} - \lambda_{34} [^3He] [^4He] \\ {d[^3He] \over dt}
&=& \lambda_{11} {[H]^2 \over 2} - 2 \lambda_{33} {[^3He]^2 \over 2} -
\lambda_{34} [^3He] [^4He] \\ {d[^4He] \over dt} &=& \lambda_{33}
{[^3He]^2 \over 2} + \lambda_{34} [^3He] [^4He] . \\
\end{eqnarray*}
After some time $^3He$ reaches equilibrium abundance:
\be
[^3He] = - \frac{1}{2} \frac{\lambda_{34}}{\lambda_{33}} [^4He] + 
\left[ \frac{1}{4} \left( \frac{\lambda_{34}}{\lambda_{33}} \right)^2 
[^4He]^2 + \frac{1}{2} \frac{  \lambda_{11}}{ \lambda_{33}} [H]^2  
\right]^{1/2}.
\ee
As the temperature increases the quantum tunneling probability that
governs the fusion of the $^3He + ^3He$ system exponentially increases
(at a rate much faster than the rate of the $pp$ reaction) 
\footnote{Sometimes this exponential dependence is converted to
dependence on a single power of temperature yielding very large powers
especially for heavier systems.}. Hence the factor
$\lambda_{11}/\lambda_{33}$ decreases quickly as we move closer to the
core of the Sun (increasing temperature) and the $^3He$ equilibrium
abundance gets to be rather small. In physical terms, at the high
temperatures of the solar core $^3He$ is burned as fast as it is
produced. Also, as one gets away from the core all nuclear reactions
cease to proceed since the temperature gets lower. Consequently there
is a region in the Sun (around 0.2 R$_{\odot}$) where the temperature
is high enough to produce $^3He$, but not high enough to burn it. In
this region a significant amount of $^3He$ can build up at
equilibrium.

Note that after the $^3$He equilibrium one has
\begin{eqnarray*}
{d[H] \over dt} &=& - \lambda_{11} [H]^2 - 2  \lambda_{34}
[^3He]_{eq}  [^4He] \\
{d[^4He] \over dt} &=& {1\over 4}\lambda_{11} [H]^2 +{1\over 2} 
\lambda_{34} [^3He]_{eq} [^4He] \\
\end{eqnarray*}
where the condition 
\be
{d[^4He] \over dt} = - {1\over 4} {d[H] \over dt}
\ee
is always satisfied. 

Three percent of the energy released from the sun is  carried away as
neutrinos. This neutrino flux can be  calculated with relatively high
precision \cite{bahc,baha} The neutrino flux predicted by the Standard
Solar Model calculations of Bahcall and Pinsonneault is shown in Figure
 1. 

\begin{figure}[t]
\vspace{8pt} \centerline{\hbox{\epsfxsize=3 in \epsfbox[41 204 500
560]{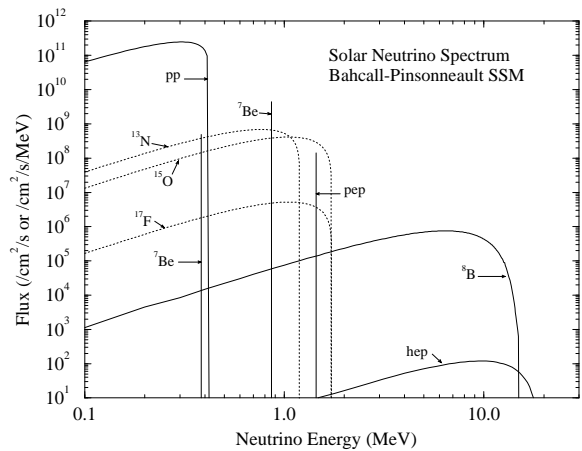}}}
\caption{
Neutrino fluxes predicted by the Bahcall and Pinsonneault.}
\vspace{8pt}
\end{figure}

\subsection{Experimental Situation}

\indent

The solar neutrino flux was first measured in the chlorine based
Homestake detector \cite{cl}. Its directionality (i.e., coming from the
sun) was established in the water Cerenkov detector Kamiokande 
\cite{kams1}. The low energy component of the neutrino flux, coming
from the $pp$ reaction (which constitutes a large component of the
flux) was observed using gallium based SAGE and GALLEX detectors
\cite{gallexnew}$^{\!-\,}$\cite{sagenew}.  
Chlorine, water, and gallium experiments are sensitive to different 
components of the solar neutrino energy spectrum. 
The observed solar neutrino
flux is deficient relative to what is predicted by the standard solar
model. A summary of the current status of the solar neutrino
experiments is given in Table 2 \cite{reviews}.  The solar neutrino
observations are not easily reconciled with the predictions of the
standard solar model. When compared to the theoretical predictions of
the neutrino flux (shown in Table 3) all experiments observe a
deficit,, the amount of which seems to depend on neutrino energy.

\begin{table}[t]
\caption{Results of solar neutrino experiments (1 SNU = 10$^{-36}$
captures per target atom per second).}
\vspace{8pt} \centering
\begin{tabular}{|c|c|c|}
\hline {\em Experiment} & {\em Threshold} & {\em Data}  \\
\hline  \hline
Homestake \cite{clnew}  &  0.814 MeV& $2.56 \pm 0.16 \pm 0.14$ SNU \\
\hline GALLEX \cite{gallexnew}  & 0.233 MeV& $70 \pm 7 \pm 4$ SNU \\
\hline SAGE \cite{sagenew}  & 0.233 MeV& $72 \pm 12 \pm 7$ SNU \\
\hline Kamiokande \cite{kamnew}  & 7.5 MeV& $(2.80 \pm 0.19 \pm
0.33)\times 10^{10}$  m$^{-2}$ s$^{-1}$.   \\ \hline Superkamiokande
\cite{superkamnew}  & 6.5 MeV & $(2.51 \pm 0.14 \pm  0.18)\times
10^{10}$  m$^{-2}$ s$^{-1}$.  \\ \hline
\end{tabular}
\vspace{8pt}
\end{table}
\begin{table}[b]
\caption{Theoretical predictions for the 
solar neutrino experiments for two different Standard Solar Models.}
\vspace{8pt} \centering
\begin{tabular}{|c|c|c|}
\hline {\em Experiment} & {\em SSM1 \cite{bp}} & {\em SSM2 \cite{tc} }
\\ \hline 
\hline Chlorine  &  $9.5 ^{+1.2}_{-1.4}$ SNU  & $6.4\pm 1.4$ SNU \\
\hline Gallium  &  $137 ^{+8}_{-7}$ SNU & $122\pm7$ SNU  \\ \hline
Water &    $ (6.62^{+0.93}_{-1.12})\times 10^{10}$m$^{-2}$ s$^{-1}$.
&  $ (4.4 \pm 1.1)\times 10^{10}$ m$^{-2}$ s$^{-1}$. \\ \hline
\end{tabular}
\vspace{8pt}
\end{table}

It is easy to demonstrate that, with the presently achieved
experimental precision, a solar neutrino problem exists independent of
any detailed model of the Sun.  Since the same nuclear reactions that
produce neutrinos also produce the rest of the solar energy, it is
possible to put constraints on the neutrino fluxes.  It takes about
$10^4$ years for photons to make it to the solar surface from the core
\cite{bahc}. If the solar core is not significantly changing over this
time scale we can write \cite{lum}$^{\!-\,}$\cite{bahlum} \be
\frac{L_{\odot}}{4\pi r^2} =\sum_i \left( Q - \langle E \rangle
\right)_i \phi_i, \ee where $r$ is the average Sun-Earth distance (1
a.u.), $Q$ is the energy released in the nuclear reactions, $\langle E
\rangle$ is the average energy loss by neutrinos, and $\phi_i$ is the
neutrino flux at earth, coming from the reaction of type $i$ (i.e., pp,
$^7$Be, $^8$B).  In a more careful treatment Eq. (25) must be
supplemented by inequalities between different fluxes to take into
account the order in which nuclear reactions take place
\cite{bahlum}. Similarly the counting rate for a given detector can be
expressed as \be S^{(J)} = \sum_i a^{(J)}_i \phi_i, \ee where
$S^{(J)}$ is the count rate at a detector of type J (i.e., chlorine,
water, gallium), and the coefficients $a^{(J)}_i$ depend on the
interaction of neutrinos in the detector, but not on the neutrino
fluxes. Currently there are three types of experiments for which an
equation like Eq. (26) can be written. If one assumes that the
neutrino fluxes are free parameters that are to be determined by the
solar neutrino experiments, using the luminosity constraint of
Eq. (25) all the existing experiments can be written in terms of two
neutrino fluxes. (For simplicity here we ignore the contribution
from the CNO neutrinos). Choosing these to be the fluxes of $^7Be$ and
$^8B$ neutrinos, one can identify allowed values in each
experiment. Such allowed regions are illustrated in Figure 2 where
experimental uncertainties are assumed to be 1$\sigma$. One observes
that, even if one arbitrarily chosen experiment is ignored, the remaining
two experiments are not compatible with any non-zero value of the
$^7Be$ neutrino flux within one sigma.

\begin{figure}[t]
\vspace{8pt} \centerline{\hbox{\epsfxsize=3 in \epsfbox[17 48 578
517]{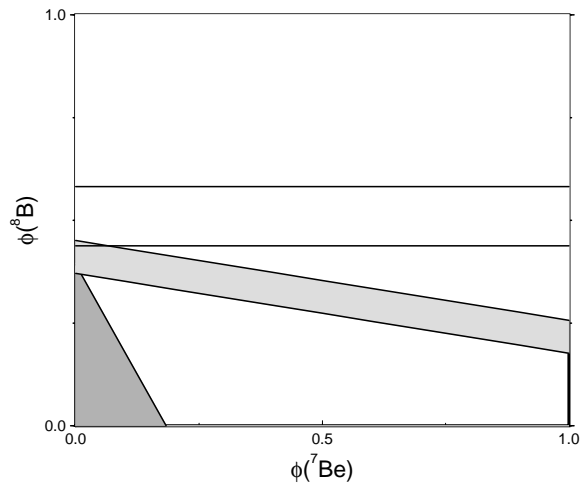}}}
\caption{
Neutrino fluxes observed in three experiments with the luminosity
constraint of Eq. (25) imposed. The regions allowed by the Kamiokande
(top), chlorine (middle) and gallium (bottom corner) are shown. For
convenience the fluxes are normalized to the SSM values of Bahcall and
Pinsonneault.}
\vspace{8pt}

\end{figure}

As we see in the next section the results from the solar neutrino
experiments imply the existence of ``new physics'', such as nonzero
neutrino masses and mixing between the different kinds of
neutrinos. Hence it is very important to test the reliability of these
experiments.  One such test is to expose these detectors to man-made
neutrino sources with a known flux and energy. Both the
GALLEX \cite{galcal} and SAGE \cite{sagecal} collaborations recently
reported results from such calibration experiments.

  There are four criteria one needs to satisfy in choosing a portable
neutrino source: (i) The mean energy of the neutrinos from the source
must be close to the mean energy of the solar neutrinos detected at
that particular detector; (ii) The activity should be such that test
measurement reaches the precision of the solar neutrino measurement;
(iii) The lifetime should be long enough so that the source can be
transported from the production site to the detector without losing
much strength; and (iv) There should be a reliable method to determine
the source activity. A $^{51}Cr$ source satisfying these criteria was
produced in Grenoble, and was used to expose the GALLEX detector. 
This was
the strongest portable neutrino source ever produced, with an activity
of $6.2 \times 10^{16}$ decays per second.  The ratio of the rate of
$^{51}Cr$ produced neutrino signals to the rate expected from the
known source activity was found to be $1.04 \pm 0.12$, providing an
overall check for the GALLEX detector and demonstrating the validity
of the basic principles of radiochemical methods to detect these rare
events (at a level of about 10 atoms per 30 tons of detector
fluid). This result, with its 11\% uncertainty, shows that the 40\%
deficit for the solar neutrino flux observed by GALLEX, as compared to
the standard solar model expectation, is unlikely to be an
experimental artifact.  The SAGE collaboration similarly utilized a
93\% enriched chromium source with an activity of $2 \times 10^{16}$
decays per second. The irradiation of the enriched chromium was
carried out at the fast neutron reactor in Aktau, Kazakhstan. At the
start of the SAGE exposure, their initial source intensity was
expected to produce about 14.7 germanium atoms a day as compared to
the solar neutrino background of 0.3 atoms a day, about 50 times
``brighter'' than the sun. Their result for the ratio of the rate of
source produced neutrino signal to the rate expected from the known
source activity is $0.93\pm 0.15 \pm 0.17$.  This result also
indicates that the deficit of solar neutrinos observed in the SAGE
experiment is unlikely to be due to some instrumental problem in the
experiment. Hence the results from GALLEX and SAGE both strongly
support the notion of a real deficit of solar neutrinos below that
predicted by the Standard Solar Models. These experiments in turn can
be used to assess the excited state contributions to the $^{71}Ga$
neutrino capture cross-section \cite{hahax}, and neutrino vacuum
oscillations between the source and the detector \cite{sourcevac}.

Earlier results from the Homestake experiment hinted a possibility of
anti-correlation between solar neutrino counts and sunspot numbers.
It is rather difficult to understand such a behavior in a theoretical
framework.  The most straightforward explanation for short-term
time-variation of the neutrino flux is to assume the existence of a
neutrino magnetic moment \cite{lam}.  However, to obtain an
anti-correlation of the observed amount, one needs rather large
magnetic moments \cite{loreti}, seemingly inconsistent with recent
bounds obtained from stellar cooling rates for plasmon decay into
neutrino-antineutrino pairs. \cite{raffeltbook} (However neutrino
magnetic moment could be important for supernova dynamics
\cite{yeniqian}).  The present status of the neutrino magnetic moment
solution to the solar neutrino problem is summarized in
Ref. 41.

Solar neutrinos are not the only experimental probes of the
sun. Information from helioseismological pulsation observations
complement information obtained by solar neutrino experiments. The
standing acoustic waves (p-waves) cause the solar surface to vibrate
with a characteristic period of about five minutes. By observing red-
and blue-shifts of patches of the solar surface, projecting them on
spherical harmonics, and finally Fourier transforming with respect to
the observation time one can obtain eigenfrequencies of the solar
p-modes with great accuracy. For very high overtones (for a
spherically-symmetric three-dimensional object such as the Sun these
are characterized by two large integers), the equations describing
p-modes simplify \cite{who} and one can reliably obtain a sound
velocity profile for the outer half of the Sun using direct Abelian 
inversion. The sound density
profile obtained this way agrees with the predictions of the standard
solar model. By studying discontinuities in the sound velocity
profile, it is also possible to reliably extract the location of the
bottom of the convective zone \cite{chris}.  More recent experimental
developments made extensive helioseismic data available, including
results from the GONG network \cite{gong} and SOI/MDI project on the
SOHO satellite \cite{soho}. These data include lower overtones that
penetrate the solar core and make a direct inversion for the solar
internal sound speed possible down to 0.05 R$_{\odot}$. Different
standard solar models are generally in good agreement with the
observations \cite{bir,iki}.

\subsection{Implications of the Data}
\indent

Over the years various astrophysical solutions to explain the
deficiency of the solar neutrinos were proposed. The model-independent
argument given in the previous subsection and sketched in Figure 2
severely limits, but not necessarily eliminates, an astrophysical
solution. One may attempt to lower the core temperature of the Sun,
but in doing so both the $^7Be$ and $^8B$ neutrino fluxes are
suppressed, contradicting the data \cite{hight}.  Recently an
alternative astrophysical solution was proposed \cite{haxtonmodel},
namely slow mixing of the solar core on time scales characteristic of
the $^3He$ equilibration discussed in Section 3.1. The basic idea here
is to find a mechanism to bring $^3He$ from 0.2 R$_{\odot}$ to the
inner core where the temperature is higher. Since after the $^7Be$
equilibrium is reached $^7Be$ equilibrium abundance follows $^3He$
abundance, $^7Be$ is then burned at the higher temperatures prevalent
at $\sim 0.05 R_{\odot}$.  A high value of the temperature does
not significantly change the rate of the electron-capture reaction
$e^-+ ^7$Be$ \rightarrow ^7$Li$ + \nu_e$, but exponentially enhances
the rate of the reaction $p+^7$Be$ \rightarrow ^8$B$ +\gamma$ where
the first step is quantum mechanical tunneling. In fact the latter
reaction is enhanced so much that it is necessary to reduce the core
temperature from the Standard Solar Model value in the simple model
Cumming and Haxton considered. Imposing such a lower temperature on
the Standard Solar Model is ruled out by the helioseismological data
\cite{iki}. Whether one can build a non-standard solar model along the
ideas of Cumming and Haxton and still be consistent with the
helioseismology needs  to be further explored. 

An alternative approach is to explain the deficiency with new
neutrino physics, either using the MSW effect discussed in Section
2.2, or using the vacuum oscillations discussed in Section 2.1. During
the last decade such analyses were given by many authors. Here we
quote the most recent such analysis given by Hata and Langacker
\cite{newhata}. The parameter space allowed by the current data is
shown in Figure 3 as calculated by them. 
The MSW solution currently seems to be the most-favored solution to
the solar neutrino problem. 

\begin{figure}[t]
\centerline{\hbox{\epsfxsize=2.5in \epsfbox[63 87 492 
654]{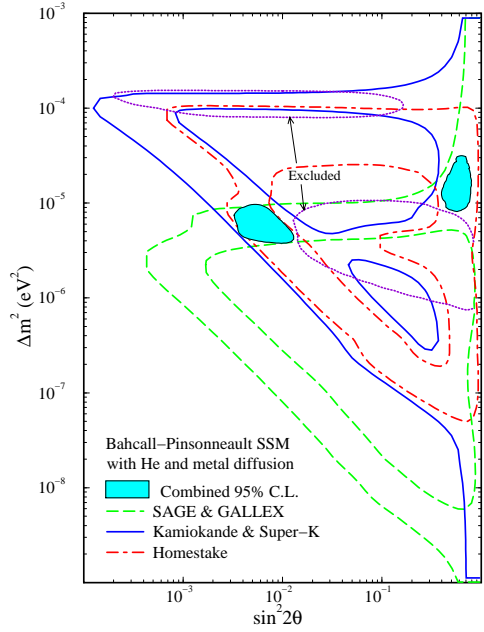}}}
\caption{MSW parameter space (shaded regions) allowed by observations 
assuming Bahcall-Pinsonneault standard solar model with He diffusion as
calculated by Hata and Langacker. Also shown are the regions excluded
by the Kamioka day-night data.}


\end{figure}

\subsection{Future Experiments}
\indent 

Neutrino oscillation solutions to the solar neutrino problem
convert electron neutrinos into neutrinos of 
other flavors. Testing the MSW
scenario requires detecting these other flavors. One experiment
designed to achieve this goal is Sudbury Neutrino Observatory (SNO).
The heavy water Cerenkov detector to be located at SNO is expected to
have a sensitivity approximately 50 times that of the Homestake
experiment \cite{sno}. The neutrinos are detected through the charged
current reaction \be \nu_e + d \rightarrow p + p + e^-, \ee and the
neutral current reactions \be \nu_e + e^- \rightarrow \nu_e + e^-, \ee
\be \nu_x (\overline{\nu}_x)+ d \rightarrow \nu_x (\overline{\nu}_x)+
p + n.  \ee The electron antineutrinos coming from a supernova or a
possible spin-flavor precession in the Sun can also be detected
through the reaction \be \overline{\nu}_e + d \rightarrow n + n +
e^+. \ee The produced neutrons will be detected either by using $(n,
\gamma )$ reactions on nuclear targets or by using $^3He$ proportional
counters.  The electrons coming from the reaction in Eq. (27) are
essentially monochromatic with energies $\sim E_{\nu} - 1.44$ MeV and
they have a very different angular distribution ($1- \cos \theta_e
/3$) with respect to the neutrino direction than that of the electrons
coming from the reaction in Eq. (28) (which are constrained to the
forward cone).  It will be possible to measure not only the total
flux, but also the energy dependence of the neutrino flux at SNO.  In
addition to detecting solar neutrinos, SNO could be a useful tool in
studying a galactic supernova \cite{snous}.

Another experiment currently under construction is BOREXINO
\cite{borexino}. which will observe neutrino electron scattering in
an organic scintillator with a threshold of 250 keV. It will have the
capability of measuring $^7Be$ neutrino flux and looking for seasonal
flux variations due to vacuum oscillations and day/night effect.
Furthermore BOREXINO has a good sensitivity to an antineutrino signal
(about 20 ev/year). As mentioned above an antineutrino signal, 
not expected by the
standard solar model, could be an indication of a neutrino magnetic
moment \cite{borexinous}.

\section{Atmospheric Neutrinos}\label{subsec:atmos}

\indent 

Atmospheric neutrinos arise from the decay of secondary pions, kaons,
and  muons produced by the collisions of primary cosmic rays with the
oxygen and nitrogen 
nuclei in the upper atmosphere. For energies less than 1
GeV all the  secondaries decay :
\begin{table}[t]
\caption{
Ratio of Ratios, R of Eq. (33), as observed in different experiments.}
\vspace{8pt} \vspace{8pt} \centering
\begin{tabular}{|c|c|}
\hline {\em Experiment} & {\em R} \\ 
\hline 
\hline  \indent Kamioka
\cite{kami}  &  $0.60 \pm 0.06 $ \\ \hline  \indent Superkamioka
\cite{superkamnew}  &  $0.67 \pm 0.05 \pm 0.06$ \\ \hline \indent IMB
\cite{imb}  & $0.54 \pm 0.05 \pm 0.07$ \\ \hline \indent Soudan 2
\cite{soudan}  & $0.72 \pm 0.19 \pm 0.07$ \\ \hline \indent Nusex
\cite{nusex}  &  $1.0 \pm 0.3$ \\ \hline \indent Frejus \cite{frejus}
&  $0.99 \pm 0.13 \pm 0.08$ \\ \hline Baksan \cite{soudan}  & $0.95
\pm 0.22 $ \\ \hline 
\end{tabular}
\end{table}

\begin{eqnarray}
\pi^{\pm} (K^{\pm}) &\rightarrow &\mu^{\pm} + \nu_{\mu}
(\overline{\nu}_{\mu}), \nonumber\\ \mu^{\pm} &\rightarrow & e^{\pm} +
\nu_e (\overline{\nu}_e) +  \overline{\nu}_{\mu} (\nu_{\mu}).
\end{eqnarray}
Consequently one expects the ratio

\begin{equation}
r = (\nu_e + \overline{\nu}_e) / (\nu_{\mu} + \overline{\nu}_{\mu})
\end{equation}

\noindent
to be approximately 0.5 in this energy range. Detailed Monte Carlo
calculations \cite{gaisser}, including the effects of muon
polarization, give $  r \sim 0.45$. Since one is evaluating a ratio of
similarly calculated processes, systematic errors are significantly
reduced. Different groups estimating  this ratio, even though they
start with neutrino fluxes which can differ in magnitude by up to
25\%, all agree within a few percent \cite{bludman}. The ratio
(observed to predicted) of ratios

\begin{equation}
R = {(\nu_{\mu} / \nu_e)_{\rm data} \over (\nu_{\mu} / \nu_e)_{\rm
Monte  Carlo} }
\end{equation}

\noindent
was determined in several experiments as summarized in Table 4. There
seems to  be a persistent discrepancy between theory and
experiment. Neutrino oscillations are
generally invoked to explain this discrepancy \cite{osc}. 

Experimentally the ratio of ratios, $R$, appears to be independent of
zenith angle. The observed zenith angle distribution of low energy
atmospheric neutrinos is consistent with no oscillations or with a
large number of oscillations for all source-detector
distances. Explanations of the low energy atmospheric neutrino anomaly
based on the oscillation of two neutrino flavors require that the
oscillating term \cite{bilen}, $\cos(\frac{\delta m^2 L}{2 E})$,
average to zero for even the shortest source-detector distances ($L<
50$ km for neutrinos from directly overhead.) For neutrinos in the
energy range 0.1 to 1 GeV this condition is satisfied for $\delta m^2
> 10 ^{-3} eV^2$.  If the atmospheric neutrino anomaly is resolved by
the oscillations of muon into tau neutrinos, this value of $\delta
m^2$ is consistent with a tau neutrino mass relevant to hot dark
matter and supernova dynamics (cf. Section 5).  
It is also possible to make a search
in the three-neutrino-flavor parameter  space and identify regions in
this parameter space compatible with the  existing atmospheric and
solar neutrino data within the vacuum oscillation  scheme \cite{andy}.

\section{Neutrino Flavor Mixing in Supernovae}

\indent

Understanding neutrino transport in a supernova is an essential part
of understanding supernova dynamics. 
Neutrino transport in a medium like supernova is a complicated 
process which needs to be treated numerically taking into account 
many different pieces of physics. 
In these lectures only the effects of neutrino flavor mixing on 
supernova dynamics are covered. 
In a core-collapse  driven supernova, the inner core
collapses subsonically, but the outer part of  the core
supersonically. At some point during the collapse, when the nuclear
equation of state stiffens, the inner part of the core bounces, but
the outer  core continues falling in. The shock wave generated at the
boundary loses its  energy as it expands by dissociating material
falling through it into free  nucleons and alpha particles. For a
large initial core mass, the shock wave  gets stalled at $\sim$ 200 to
500 km away from the center of the proto-neutron  star
\cite{mayle}. Meanwhile, the proto-neutron star, shrinking under its
own  gravity, loses energy by emitting neutrinos, which only interact
weakly and  can leak out on a relatively long diffusion time
scale. The question to be investigated then is  the possibility of
regenerating the shock by neutrino heating.

\begin{figure}[t]
\vspace{8pt} \centerline{\hbox{\epsfxsize=3 in \epsfbox[105 242 459
518]{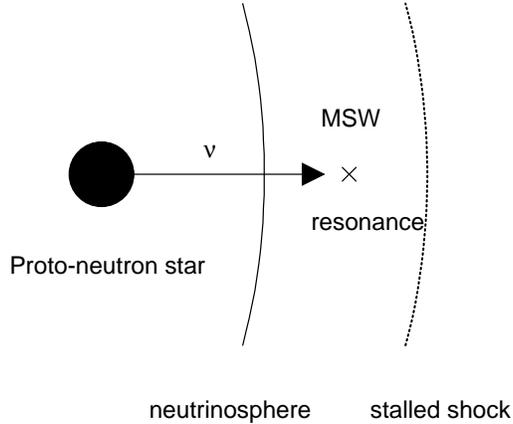}}}
\caption{
Neutrinosphere and the stalled shock in a core-collapse driven
supernova.}  \vspace{8pt}
\end{figure}

The situation at the onset of neutrino heating is depicted in Figure
4. The  density at the neutrinosphere is $\sim 10^{12}$g cm$^{-3}$
and the density at the position of the stalled shock is \cite{mayle}
$\sim 2 \times 10^7$ g cm$^{-3}$. Writing the MSW resonance density in
appropriate units:

\begin{equation}
\rho_{\rm res} = 1.31 \times 10^7 \left( {\delta m^2 \over {\rm eV}^2}
\right)  \left( { {\rm MeV} \over E_{\nu} } \right) \left( {0.5 \over
{\rm Y}_e}  \right) {\rm g} \> {\rm cm}^{-3},
\end{equation}

\noindent
one sees that, for small mixing angles, $E_{\nu} \sim 10$ MeV, and
cosmologically interesting $\delta m^2 \sim 1 - 10^4$ eV$^2$, there is
an MSW resonance point between the neutrinosphere and the stalled
shock. 

Most neutrinos emitted from the core are produced by a neutral current
process, and so the luminosities are approximately the same for all
flavors.  The energy spectra are approximately Fermi-Dirac with a zero
chemical potential characterized by a neutrinosphere temperature. The
$\nu_{\tau}, {\overline \nu}_{\tau}, \nu_{\mu}, {\overline \nu}_{\mu}$
interact with matter only via neutral current interactions. These
decouple at relatively small radius and end up with somewhat high
temperatures, about 8 MeV. The ${\overline \nu}_e$'s decouple at a
larger radius because of the additional charged current interactions
with the protons, and consequently have a somewhat lower temperature,
about 5 MeV. Finally, since they undergo charged current interactions
with more abundant neutrons, $\nu_e$'s decouple at the largest radius
and end up with the lowest temperature, about 3.5 to 4 MeV.  An MSW
resonance between the neutrinosphere and the stalled shock can then
transform $\nu_{\tau} \leftrightarrow \nu_e$, cooling $\nu_{\tau}$'s,
but heating $\nu_e$'s. Since the interaction cross section of electron
neutrinos with the matter in the stalled shock increases with
increasing energy, it may be possible to regenerate the shock. Fuller
{\it et al.} found that for small mixing angles between $\nu_{\tau}$
and $\nu_e$ one can get a 60\% increase in the explosion energy
\cite{mayle}.

There is another implication of the $\nu_{\tau}$ and $\nu_e$ mixing in
the supernovae. Supernovae are possible r-process sites \cite{burb},
which requires a neutron-rich environment, i.e., the ratio of
electrons to baryons, $Y_e$, should be less than one half. $Y_e$ in
the nucleosynthesis region is given approximately by \cite{qian}

\begin{equation}
Y_e \simeq {1 \over 1+ \lambda_{{\overline \nu}_e p} / \lambda_{ \nu_e
n}}  \simeq {1 \over 1 + T_{{\overline \nu}_e} / T_{ \nu_e}}, 
\end{equation}

\noindent
where $\lambda_{ \nu_e n}$, etc. are the capture rates. Hence if
$T_{{\overline \nu}_e} > T_{\nu_e}$, then the medium is
neutron-rich. As we discussed above, without matter-enhanced neutrino
oscillations, the neutrino temperatures satisfy the inequality $T_{
\nu_{\tau}} >T_{{\overline \nu}_e} > T_{ \nu_e}$. But the MSW effect,
by heating $\nu_e$ and cooling $\nu_{\tau}$ can reverse the direction
of inequality, making the medium proton-rich instead. Hence the
existence of neutrino mass and mixings puts severe constraints on
heavy-element nucleosynthesis in supernova. These constraints are
investigated in Ref. 67. 

\section{Neutrino Propagation in Stochastic Media}

\indent

In implementing the MSW solution to the solar neutrino problem one
typically assumes that the electron density of the Sun is a
monotonically decreasing function of the distance from the core and
ignores potentially de-cohering effects \cite{sawyer}. To explore 
the validity of these assumptions 
parametric changes in the density and matter currents were considered
\cite{othernoise}. More recently Loreti and Balantekin
\cite{orignoise} considered neutrino propagation in stochastic
media. They studied the situation where the electron density in the
medium has two components, one average component given by the Standard
Solar Model or Supernova Model, etc. and one fluctuating
component. Then the Hamiltonian in Eq. (7) takes the form \be \hat H =
\left({{-\delta m^2}\over 4E} \cos 2\theta + {1\over \sqrt{2}}
G_F(N_e(r) + N^r_e(r))\right){\sigma_z} + \left({{\delta m^2}\over 4E}
\sin 2\theta \right) {\sigma_x}.  \ee where one imposes for
consistency \be \langle N^r_e(r)\rangle = 0, \ee and a two-body
correlation function \be \langle N^r_e(r)N^r_e(r^{\prime}) \rangle =
{\beta}^2 \ N_e(r) \ N_e(r^{\prime}) \ \exp(-|r-r^{\prime}|/\tau_c), 
\ee 
where $\beta$ is a measure of the amount of the fluctuation, and 
$\tau_c$ is the correlation length. 
In the calculations of the Wisconsin group the fluctuations are
typically taken to be subject to colored noise, i.e., higher order
correlations \be f_{12 \cdots }=\langle N^r_e(r_1)N^r_e(r_2) \cdots
\rangle\ee are taken to be 
zero for an odd number of $N^r_e$'s and 
\be f_{1234}= f_{12}f_{34} + f_{13}f_{24} +
f_{14}f_{23},\ee and its appropriate generalization for an even number 
of $N^r_e$'s. 

\begin{figure}[t]
\vspace{8pt} \centerline{\rotate[r]{\epsfxsize=3in
\epsfbox{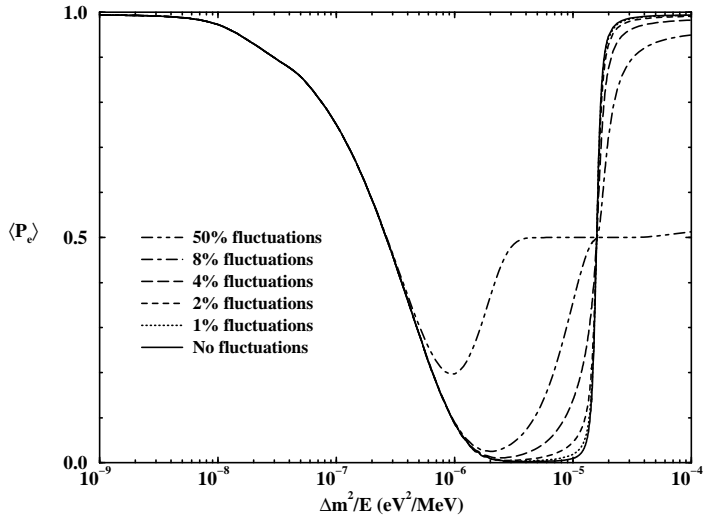}}} 
\caption{
Mean electron neutrino survival probability in the sun with
fluctuations. The average electron density is given by the Standard
Solar Model of Bahcall and Pinsonneault and $\sin^2 2 \theta=0.01$.} 
\vspace{8pt}
\end{figure}

Mean survival probability for the electron neutrino in the Sun is
shown in Figure 5  where fluctuations \cite{newnoise} are imposed on
the average solar electron density given by the Bahcall-Pinsonneault
model. In this figure there are two salient features. The first one is
that for very large fluctuations the MSW effect is
``undone''. Complete flavor de-polarization is achieved, i.e. the
neutrino survival probability is 0.5, the same as the vacuum
oscillation probability for long distances. To illustrate this
behavior the results from the physically unrealistic case of 50\%
fluctuations are shown. The second one is that the effect is largest
when the neutrino propagation in the absence of fluctuations is
adiabatic. This immediately suggests implementing this scenario to the
neutrino conversion in a
core-collapse supernova \cite{supernoise} as described in Section 5.  
These conclusions were also confirmed by other authors. 
\cite{nunokawa}$^{\!-\,}$\cite{burgess}
Propagation of a neutrino with a magnetic moment in a random magnetic 
moment has also been
investigated. \cite{orignoise}$^{,}$\cite{ranmagnetic} 

\begin{figure}[t]
\vspace{8pt} \centerline{\rotate[r]{\epsfxsize=3in
\epsfbox{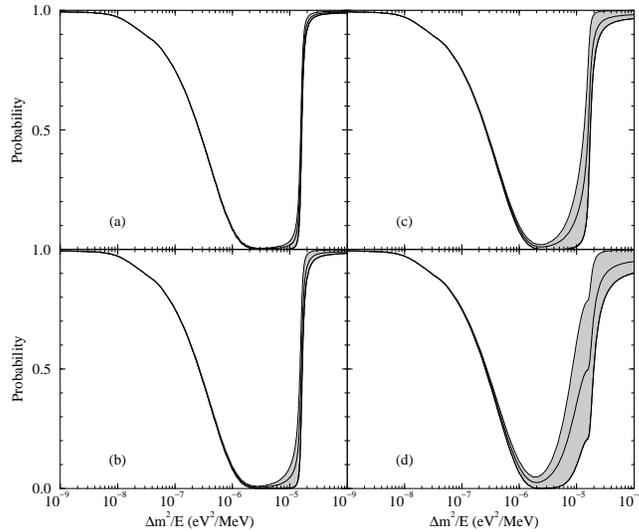}}} 
\caption{ Mean electron neutrino survival probability plus minus
$\sigma$ in the sun with fluctuations. The average electron density is
given by the Standard Solar Model of Bahcall and Pinsonneault and  
$\sin^2 2 \theta=0.01$. Panels (a), (b), (c), and (d) correspond to an
average fluctuation of 1\%, 2\%, 4\%, and 8\% respectively. 
}
\vspace{8pt}
\end{figure}

Using Eqs. (36) through (40), it is possible to calculate not only the
mean survival probability, but also the variance, $\sigma$, of the
fluctuations to get a feeling for the distribution of the survival
probabilities. Results obtained for several representative cases are
shown \cite{newnoise} in Figure 6. This broadening of the survival 
probabilities may be measurable in real-time experiments sensitive to 
the low-energy component of the solar neutrino flux. 

In these calculations the correlation length $\tau_c$ is taken to be
very small, of the order of 10 km., to be consistent with the
helioseismic deduction of the sound speed \cite{iki}. In the
opposite limit of very large correlation lengths are very interesting
result is obtained \cite{supernoise}, namely the averaged density
matrix is given as an integral 
\be
\lim_{\tau_c\to\infty}\langle\hat \rho(r)\rangle =
{1\over{\sqrt{2\pi \beta^2}}} \int_{-\infty}^{\infty} dx
\exp[{-x^2/(2\beta^2)}]
\hat \rho(r,x),
\ee
reminiscent of the channel-coupling problem of nuclear
physics in the sudden limit. \cite{takigawa} 

Finally we should mention that if the magnetic field in a polarized
medium has a domain structure with different strength and direction
in different domains, the modification of the potential felt by the
neutrinos due polarized electrons will have a random
character as depicted in Eq. (39) \cite{polarized}. 

\section*{Acknowledgments}

\indent

This research was supported in part by the U.S. National Science
Foundation Grant No.\ PHY-9605140 at the University of Wisconsin, and
in part by the University of Wisconsin Research Committee with funds
granted by the Wisconsin Alumni Research Foundation.

\end{document}